\documentstyle[preprint,aps]{revtex}
\begin{document}
\newif\ifplot
\plottrue
%\plotfalse
\newcommand{\RR}[1]{[#1]}
\newcommand{\intsum}{\sum \kern -15pt \int}
\newfont{\Yfont}{cmti10 scaled 2074}
\newcommand{\Y}{\hbox{{\Yfont y}\phantom.}}
\def\O{{\cal O}}
\newcommand{\bra}[1]{\left< #1 \right| }
\newcommand{\braa}[1]{\left. \left< #1 \right| \right| }
\def\Bra#1#2{{\mbox{\vphantom{$\left< #2 \right|$}}}_{#1}
\kern -2.5pt \left< #2 \right| }
\def\Braa#1#2{{\mbox{\vphantom{$\left< #2 \right|$}}}_{#1}
\kern -2.5pt \left. \left< #2 \right| \right| }
\newcommand{\ket}[1]{\left| #1 \right> }
\newcommand{\kett}[1]{\left| \left| #1 \right> \right.}
\newcommand{\scal}[2]{\left< #1 \left| \mbox{\vphantom{$\left< #1 #2 \right|$}}
\right. #2 \right> }
\def\Scal#1#2#3{{\mbox{\vphantom{$\left<#2#3\right|$}}}_{#1}
%\kern -2pt
{\left< #2 \left| \mbox{\vphantom{$\left<#2#3\right|$}}
\right. #3 \right> }}
\draft
%\preprint{***}
\title{
The cross section minima in elastic Nd scattering: a ``smoking gun'' for three 
nucleon force effects
}
\author{H. Wita\l a$^*$, W. Gl\"ockle$^\dagger$, D. H\"uber$^\ddagger$,
J. Golak$^*$, 
H. Kamada$^{\dagger}$
}
\address{
$^\dagger$Institut f\"ur theoretische Physik II, Ruhr-Universit\"at Bochum,
D-44780 Bochum, Germany
}
\address{$^{*}$ Institute of Physics, Jagellonian University, 
PL- 30059 Cracow, Poland}
\address{$^\ddagger$Los Alamos National 
Laboratory, M.S. B283, Los Alamos, NM 87545, USA}

\date{\today}
\maketitle
\widetext
\begin{abstract}
Neutron-deuteron elastic scattering cross sections are calculated at 
different energies using modern nucleon-nucleon 
 interactions and the Tucson-Melbourne 
three-nucleon force adjusted to the triton binding energy. Predictions 
based on NN forces only underestimate nucleon-deuteron data in the minima 
at higher energies starting around 60 MeV. Adding the three-nucleon 
forces fills up those 
minima and reduces the discrepancies significantly.
\end{abstract}

\pagebreak
%\tableofcontents
%\pagebreak
\narrowtext

%\section{Introduction}

Substantial progress has been made recently in the study of the three-nucleon 
(3N) system both experimentally and theoretically. The set of data is being 
significantly enriched for cross sections and spin observables in 
elastic neutron-deuteron (nd) and proton-deuteron (pd) scattering 
and in the 3N breakup process. Theoretical formulations and numerical 
algorithms have been significantly improved with the result that 3N bound 
and scattering states can be solved exactly. In addition the nucleon-nucleon 
(NN) system is still very intensively investigated and the increased data 
set provides a sound foundation for reliable modern phase-shift 
analysis~\cite{1}. Based on these phases modern NN forces have been 
constructed by different groups~\cite{2,3,4}. These interactions 
reproduce the NN data set with unprecedented accuracy as measured by a 
${\chi}^2$/datum very close to 1. Although those forces are not yet 
linked to the underlying QCD due to well known reasons, they cover 
a wide spectrum of expected properties. There are local, weakly nonlocal 
and strongly nonlocal versions, purely phenomenological ones and 
semi-phenomenological ones guided by meson theory. They form an interesting 
basis to study few-nucleon systems. Thus theoretical tools and data are 
available to approach the basic questions about the dynamics of three 
interacting nucleons: is the unperturbed free two-nucleon (2N) interaction 
predominant or is it significantly modified due to the presence of an 
additional nucleon? Such a modification would be one out of many other 
possible three-nucleon force (3NF) mechanisms. What is their significance 
altogether? If the free NN force is sufficient to a very large extent, 
 are specific NN interactions 
 favored in the description of the 3N system? In the 
future QCD should provide theoretically consistent NN and 3N forces and 
specifically the relative importance of the latter ones for binding 
energies and scattering matrices. First steps on that ground are 
being done in chiral perturbation theory~\cite{5}. Despite of that 
still restricted 
 theoretical insight from QCD one can go ahead and compare the 
theoretical predictions obtained with modern NN interactions and model 
3N forces to experimental 3N data. There might be a clearcut signal coming 
from certain observables which deny to be explained by 3N Hamiltonians 
based on modern NN interactions only. Such a ``smoking gun'' observable 
would then put limits on present day 3N force models and would 
be also of great importance to test the future QCD based dynamics. 

The three-nucleon binding energy by itself is a first signature. The modern 
NN interactions underbind $^3H$, but to a different extent~\cite{6}. The 
essentially local ones lack binding energy of about 800 keV out of 
8.48 MeV whereas the nonlocal CD Bonn interaction~\cite{4} underbinds 
only by $\approx$500 keV. That nonlocality is generated by the Dirac 
structure of the spin 1/2 nucleon. One might conjecture that more 
dynamical nonlocalities caused by the composite nature of the nucleons 
might have additional effects, which are not sufficiently well mimicked 
by the variety of present day NN force models. Anyhow, that information 
from  $^3H$  on insufficient dynamics based on present day NN forces only 
 should be enriched by further evidence from the 3N 
continuum. 

Such a search for 3N continuum observables, which could 
serve as a ``smoking gun'' for 3NF effects has been persued ever since 3N 
continuum calculations have become feasible~\cite{7}. With the advent 
of the optimally tuned NN forces and the feasibility to also include 
3NF's into 3N continuum calculations the conclusive power of such 
calculations has increased tremendously. It is the aim of this article 
to point to such a ``smoking gun'' in the 3N continuum  based on modern 
3N Faddeev calculation. 

Before coming to that let us briefly describe the situation in 3N continuum 
studies. A detailed overview has been given recently~\cite{8}. The bulk 
of 3N scattering observables below about 100 MeV nucleon lab energy can 
be described quite well in the NN force picture only. A beautiful 
example is the total nd cross section~\cite{9}. This most simple picture 
is also quite stable in the sense that the most modern phase-equivalent 
NN force models yield essentially the same predictions. But there are 
exceptions, ``time dependent ones'', which were removed by subsequent 
measurements~\cite{10}, but more important true ones, where the data 
are reconfirmed by independent measurements. Such a distinguished 
case is the low energy vector analyzing power $A_y$ in elastic 
Nd scattering~\cite{11}. A drastic discrepancy between the predictions 
based on NN forces only and both nd and pd data has been found. Present 
day 3NF models have insignificant effects and do not remove that 
discrepancy.  
It is known, that $A_y$ depends 
very sensitively on the $^3P_j$ NN forces. Thus a trivial explanation 
might be that the $^3P$ NN phase-shift parameters from modern phase-shift 
analysis have not been settled to the true ones~\cite{12}. Presently it is 
an unsolved puzzle. If the reason does not lie in the NN forces a 3NF of still 
unknown properties will be responsible.

Another possible signature for 3NF effects is the space star configuration 
in the 3N breakup process at 13 MeV~\cite{10}. In that configuration the 
three nucleons emerge from the interaction region with equal energies 
and interparticle angles of 120$^o$ in the 3N c.m. system. Also the 
plane spanned by the three momenta 
is perpendicular to the beam direction. Two nd measurements 
agree essentially with each other but deviate from theory (in the NN 
picture only  and including 3NF models). The situation poses even more 
questions since pd data deviate very severely from the nd data pointing 
to unexpectedly large Coulomb force effects~\cite{13}. 

In the present study we investigate the angular distribution in elastic 
Nd scattering, the most simple elastic Nd scattering  observable.
The transition amplitude for this process is composed of the nucleon 
exchange part ($PG_0^{-1}$), the direct action of a 3NF and a part 
having its origin in the multiple interactions of 3 nucleons through 
2N and 3N forces:
\begin{eqnarray}
<\phi'\vert U \vert\phi> &=& <\phi'{\vert} PG_0^{-1} +  
 V_4^{(1)}(1+P) + P\tilde T + V_4^{(1)}(1+P)G_0\tilde T{\vert}{\phi}>
\label{eq1}
\end{eqnarray}
That rescattering part is expressed in terms of a $\tilde T$ operator which 
sums up all multiple scattering contributions through the integral 
equation~\cite{Actap}
\begin{eqnarray}
\tilde T{\vert}{\phi}> &=& tP{\vert}{\phi}> + 
(1+tG_0)V_4^{(1)}(1+P){\vert}{\phi}> 
+ tPG_0\tilde T{\vert}{\phi}> \nonumber \\ 
 &+& (1+tG_0)V_4^{(1)}(1+P)G_0\tilde T{\vert}{\phi}>
\label{eq2}
\end{eqnarray}
Here $G_0$ is the free 3N propagator, t the NN t-matrix, and P the sum 
of a cyclical and anticyclical permutation of three objects. The 3NF $V_4$ 
is split into 3 parts  
\begin{equation}
V_{4} = \sum_{i=1}^{3} V_4^{(i)} 
\label{eq3}
\end{equation}
where each one is symmetrical under exchange of two particles. For 
the $\pi-\pi$ exchange 3NF for instance~\cite{14} this corresponds 
to the three possible choices of the nucleon, which undergoes the (off-shell) 
$\pi-N$ scattering. The asymptotic state ${\vert}{\phi}>$ 
(${\vert}{\phi}'>$) is a product of the deuteron wave function and the 
momentum eigenstate  of the third particle. 

The exchange part comprises two processes where the incoming nucleon ends up 
as a constituent of the final deuteron and the constituents of the initial 
deuteron are free in the final state. Due to the nature of this term 
its contribution to the elastic scattering cross section is peaked at 
backward angles. The contribution from the driving term  
 $tP{\vert}{\phi}>$ and the rescattering terms in t (NN force 
contributions only) are peaked at forward angles. Therefore the 
elastic scattering  cross section exhibits a characteristic minimum 
in the angular range where the contributions of the exchange and the 
rescattering terms are of comparable order and both are small. 
This angular range around the minimum could thus be a place where the 3NF 
signal, if sufficiently strong, should appear. It would happen at those 
energies where the pure 3NF contribution to the elastic 
scattering amplitude in that minimum is comparable or larger than 
the contributions of the exchange part and the pure 2N rescattering terms. 

The pure 3NF  contribution to the transition operator U results from 
 Eqs.~(\ref{eq1}) and ~(\ref{eq2}) when only the 3NF is active:
\begin{eqnarray}
 U^{3NF}  &=&  P\tilde T^{3NF} + 
 V_4^{(1)}(1+P) + V_4^{(1)}(1+P)G_0\tilde T^{3NF}
\label{eq4}
\end{eqnarray}
with
\begin{eqnarray}
  \tilde T^{3NF}\vert\phi>  &=& V_4^{(1)}(1+P)\vert\phi> 
+ V_4^{(1)}(1+P)G_0\tilde T^{3NF}\vert\phi>
\label{eq5}
\end{eqnarray}
We expect that the contribution of $U^{3NF}$ alone is uniformly distributed 
over all angles. 

In order to check these expectations we solved  Eqs.~(\ref{eq2}) 
and ~(\ref{eq5}) 
 at the nucleon laboratory energies of 12, 65, 140 and 200 MeV using the 
modern NN interactions: AV18~\cite{3}, CD Bonn~\cite{4}, 
Nijm I and Nijm II~\cite{2}. As the 3NF we took the $2\pi$-exchange 
Tucson-Melbourne (TM) model~\cite{14} where the strong cut-off 
parameter $\Lambda$ has been adjusted individually together with each 
NN force to the experimental triton binding~\cite{6}. In the 
calculations including 3NF's all partial wave states with total 
angular momenta in the two-nucleon subsystem up to $j_{max}=3$ 
were taken into account. It is the most extensive calculation with 
3NF's in the continuum which we can perform presently. At the higher energies 
they are not fully converged with respect to $j_{max}$. The importance 
of partial waves with higher two-nucleon angular momenta are illustrated 
in fully converged solutions in the case when only 2N forces are active. 
Then we included all states up to $j_{max}=5$. Our theoretical results 
are shown in Figs.1-4 in comparison to data. Our 
theory does not include the pp Coulomb force. Therefore we should compare 
to nd data. This is only possible at rather low energies, where nd data exist 
and which agree perfectly with NN force predictions only~\cite{8}. The 
pd data also existing there agree with the nd data except at very forward 
angles, where Rutherford scattering has to show up. That interference 
with Rutherford scattering can clearly be seen in 
Figs.1 and 2 at forward angles, where the data bend 
towards smaller values. Aside from that there is a very good agreement  
at 12 MeV with theory. 
This together with the smallness of the Coulomb force effects on the 
elastic scattering cross section in the region of its minimum, as shown 
by exact calculations under the deuteron breakup threshold~\cite{Ki95},  
 supports the conjecture that a comparison of nd theory 
with pd data at even higher energies makes sense. 
The Figs.1-4 show the expected result, that  the pure 3NF contribution 
is essentially uniform in its angular dependence and we see that at 12 MeV 
it is totally negligible. 
At 65 MeV there are also a few nd data~\cite{65nd} and 
as shown in Fig.2 they come close to NN force predictions only, 
whereas the pd data~\cite{65pd} deviate strongly in the minimum. 
Without a rigorous calculation including the pp Coulomb force it has 
to remain an open question whether the deviation between the pd data 
and the NN force predictions only is due to our neglect of Coulomb 
forces in the theoretical calculations. 
On the other hand the nd data of Fig.3are 
compatible with pd data in this energy range 
and indicate only small Coulomb force effects corresponding to our conjecture. 
 Apparently precise nd data 
in the angular range of the minima for 65 MeV and higher would be highly 
desireable. Independent of that important issue we can go ahead and display 
possible 3NF effects in these minima. The discrepancy of the theory 
based on NN forces only to the pd data increases with energy as seen in 
Figs.2-4. Higher angular momentum states do not cure 
that discrepancy. They have a significant contribution, however, to the 
cross section at the higher energies at forward angles ~\cite{17}, as
 seen especially in Figs.3 and 4. As expected the pure 
3NF contribution remains essentially uniform also at the higher energies. 
With increasing energy, however, this contribution becomes significant 
in relation to the minimum value of the cross section. Being totally 
negligeable at 12 MeV it overshoots the minimum value by a factor of 
$\approx 6$ at 200 MeV. At 65 MeV the 3NF signal becomes sufficiently 
large to be seen in the minimum region. Indeed as shown in 
 Figs.2-4, including the 3NF in addition to the 2N 
interactions in the 3N Hamiltonian removes a large part of the 
discrepancy in the cross section minimum at the higher energies. Though the 
calculations with 3NF are not yet fully converged because of presently 
existing computer limitations we consider that filling of the minima as 
a ``smoking gun'' for 3NF effects. Very precise data, both in the nd and pd 
system, would therefore be highly valuable. 

We have to expect additional modifications, especially at the highest 
energies, due to relativistic effects, which have not been taken into 
account in our calculation. First estimates just based on kinematical 
factors indicate indeed a small shift of all angular distribution 
at higher energies toward 
higher values.

Finally we want to emphasize that our 
conclusions do not depend on the particular NN interaction used. As 
shown in Fig.5 taking different modern NN interactions and the 
corresponding TM 3NF leads to practically the same results. 

Summarizing, we have shown that the minima of the elastic Nd 
scattering cross sections are probably a ``smoking gun'' for 3NF effects. 
A large part of the discrepancy between modern NN potential predictions 
and data in this angular range can be removed when the TM 3NF properly 
adjusted to the triton binding is included in the 3N Hamiltonian. 
In order to check more accurately this conclusion precise Nd elastic 
scattering data at different energies in the region of the cross section 
minima are required. The optimal data would be in the nd system 
to avoid the theoretical uncertainty of pp Coulomb force effects.

\begin{center}
{\bf Acknowledgements}
\end{center}

This work was supported by the Deutsche Forschungsgemeinschaft 
 under Project No. Gl87/24-1. 
 The work of D.H. was partially supported by the Deutsche 
   Forschungsgemeinschaft under Project No. Hu 746/1-2 and
   partially by the U.S. Departement of Energy.
 The numerical calculations have been performed on the 
CRAY T90 and the CRAY T3E of the H\"ochstleistungsrechenzentrum in J\"ulich,
Germany.

\newpage
\subsection*{Figure Captions}
\begin{description}

\item{Fig.1}
{The differential Nd cross section at $E_N^{lab}=12$ MeV. 
The prediction of the CD Bonn NN interaction without (short dashed) and with 
3NF (solid curve) is compared to pd data (circles from ~\cite{12pdn1} 
and crosses from ~\cite{12pdn2}). The long dashed curve is the pure 3NF 
prediction. All the calculations are truncated at $j_{max}=3$.
}

\item{Fig.2}
{The differential Nd cross section at $E_N^{lab}=65$ MeV. 
The prediction of the CD Bonn NN interaction for  $j_{max}=3$ (short dashed 
curve) 
and  $j_{max}=5$ (long dashed curve) 
are  compared to 64.5 MeV pd data 
(circles from ~\cite{65pd}) and nd data  
 (crosses from ~\cite{65nd}). 
The CD Bonn calculation including the 3NF for  $j_{max}=3$ fills the 
minimum (solid curve). The pure 3NF 
prediction is shown as intermediately long  dashed curve.
}

\item{Fig.3}
{The differential Nd cross section at $E_N^{lab}=140$ MeV. 
Curve descritpions as in Fig.2. The pd data are: 145.5 MeV (circles) from 
~\cite{140pdn2}  and 146 MeV (crosses) from ~\cite{140pdn1}. 
The triangles are 152 MeV nd data from ~\cite{140pdn3}.
}

\item{Fig.4}
{The differential Nd cross section at $E_N^{lab}=200$ MeV. 
Curve descritpions as in Fig.2. The pd data are: 198 MeV (circles) from 
~\cite{200pd}, 200 MeV (crosses) from ~\cite{17}, 181 MeV 
(triangles) from ~\cite{140pdn2} and  216.5 MeV 
(x) from ~\cite{140pdn2}.
}

\item{Fig.5}
{The theoretical differential nd cross sections at 
$E_N^{lab}=200$ MeV  for  $j_{max}=3$. The lower group of curves in the 
minimum are: AV18 (short dashed), CD Bonn (solid), NijmI (intermediately 
long dashed) 
and Nijm II (long dashed); 
the upper group of curves in the minimum are the corresponding 
predictions with TM 3NF included.
 This demonstrates the independence from the specific choice of the NN force.
}

\end {description}

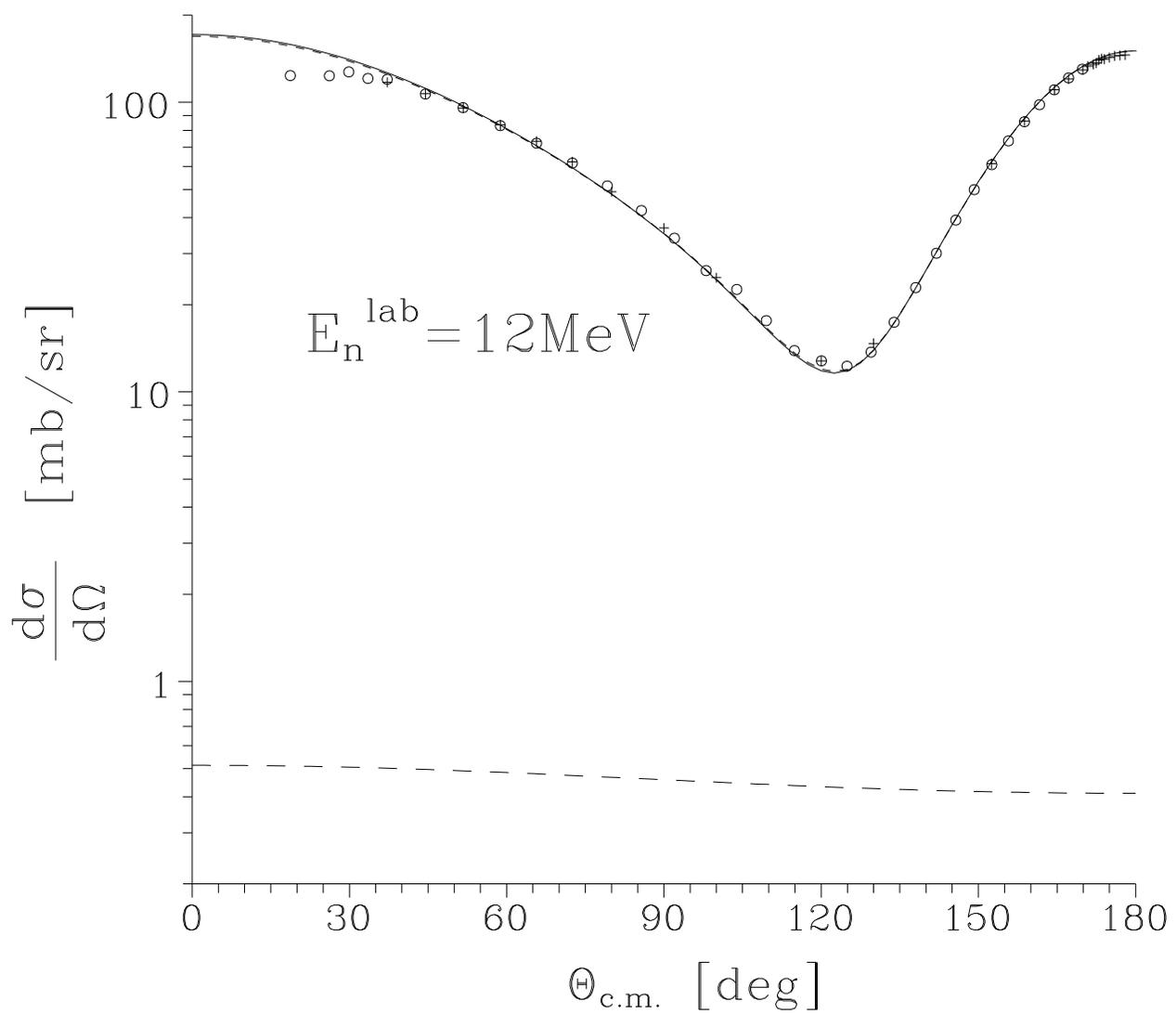
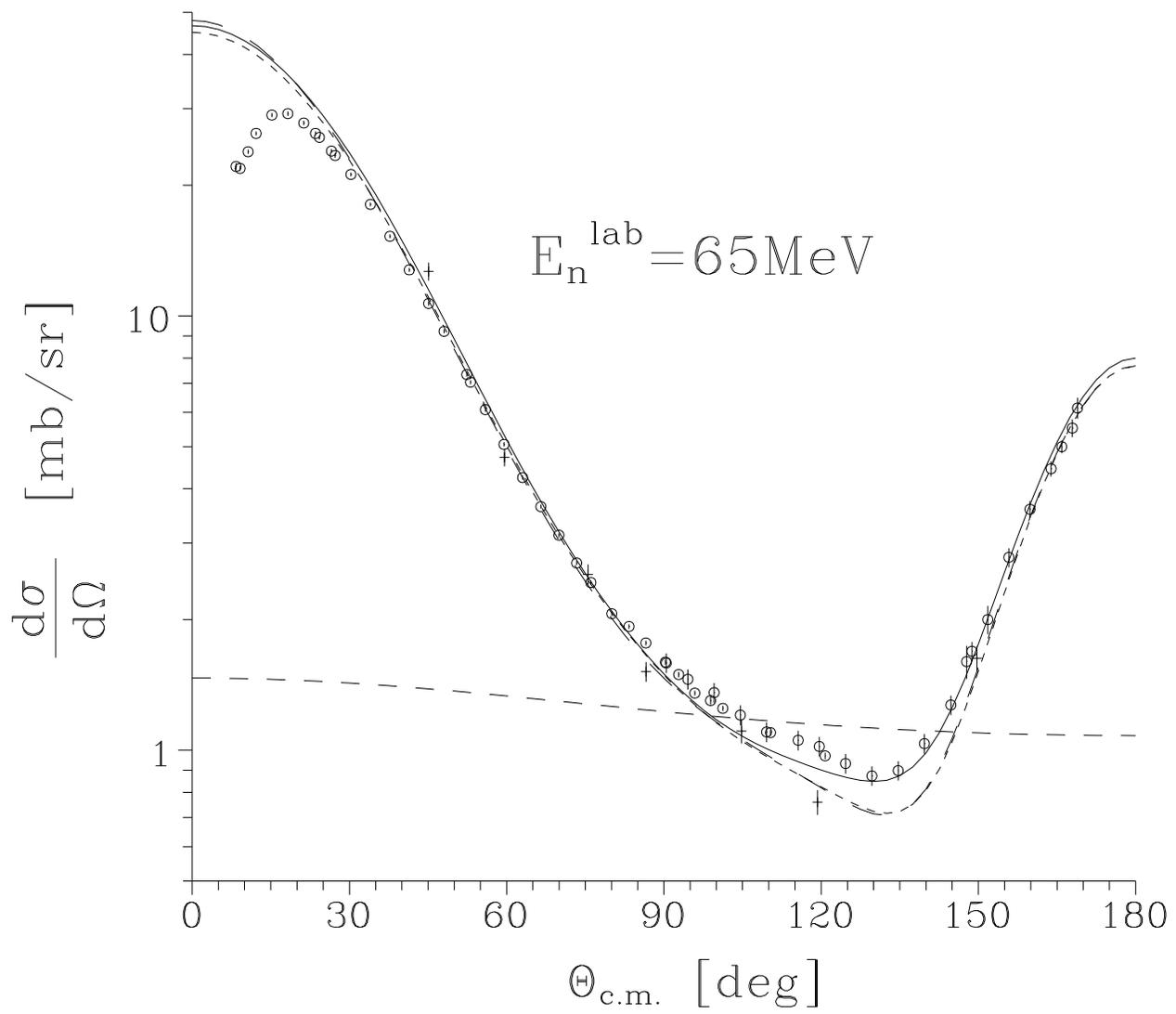
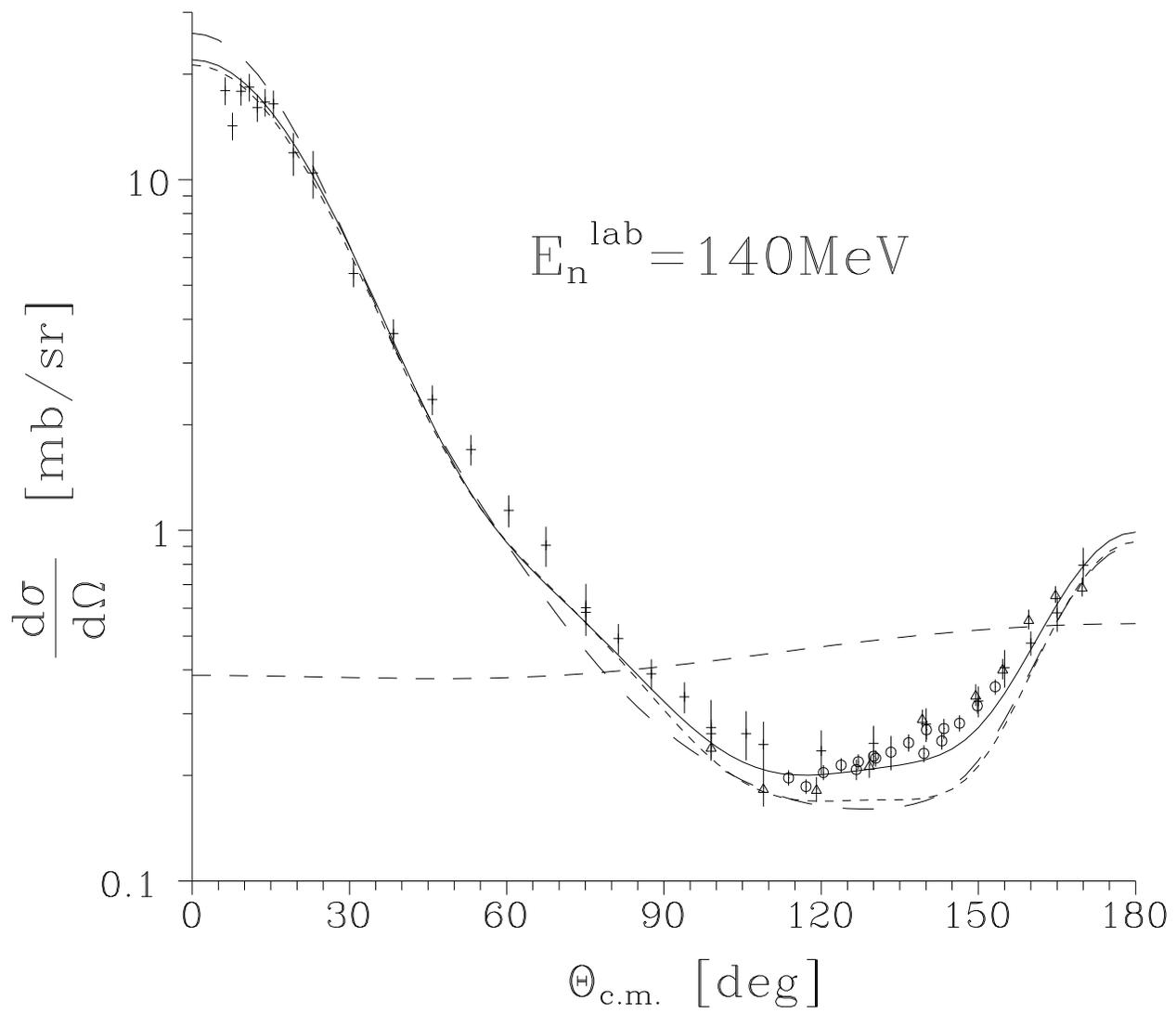
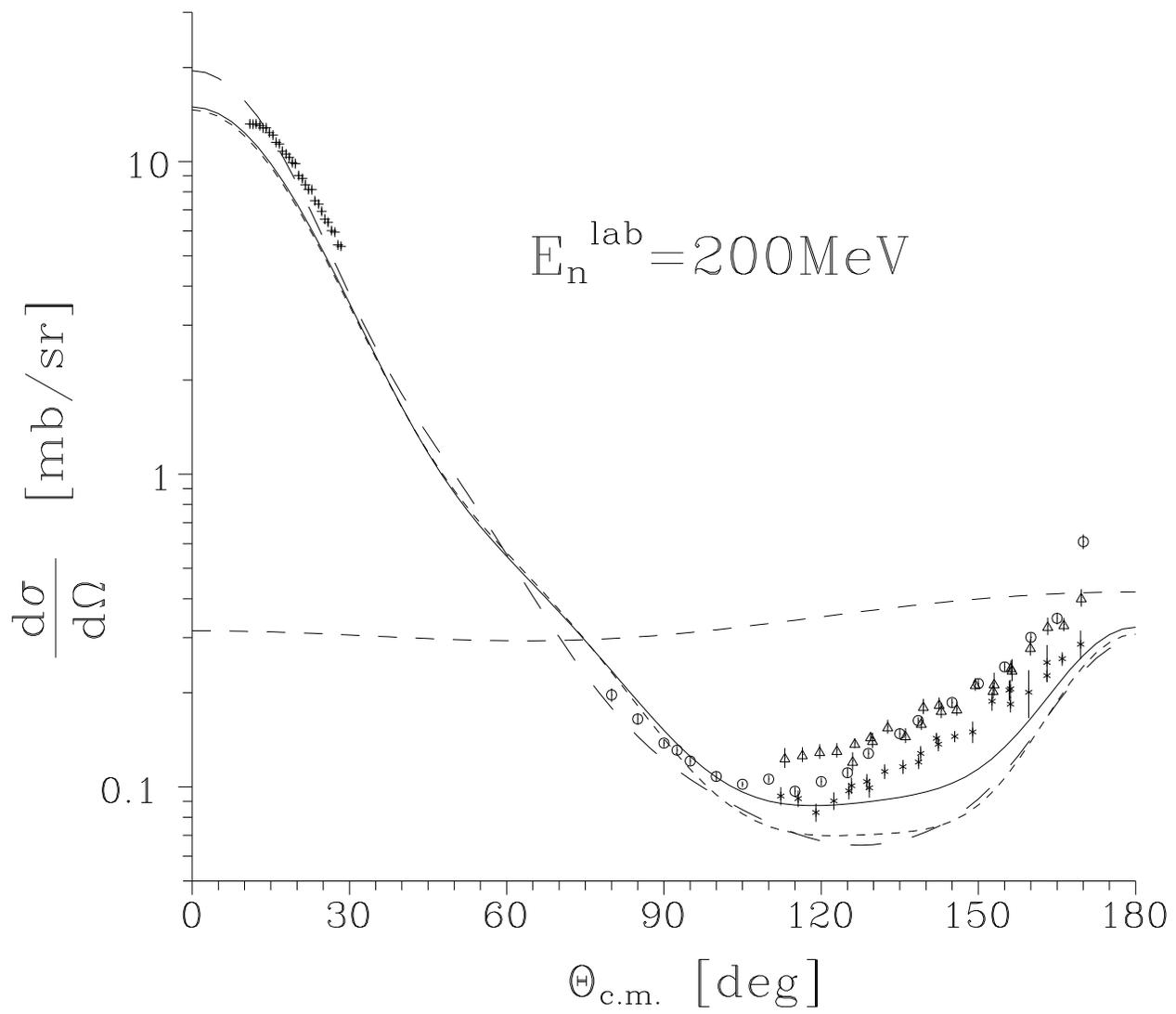
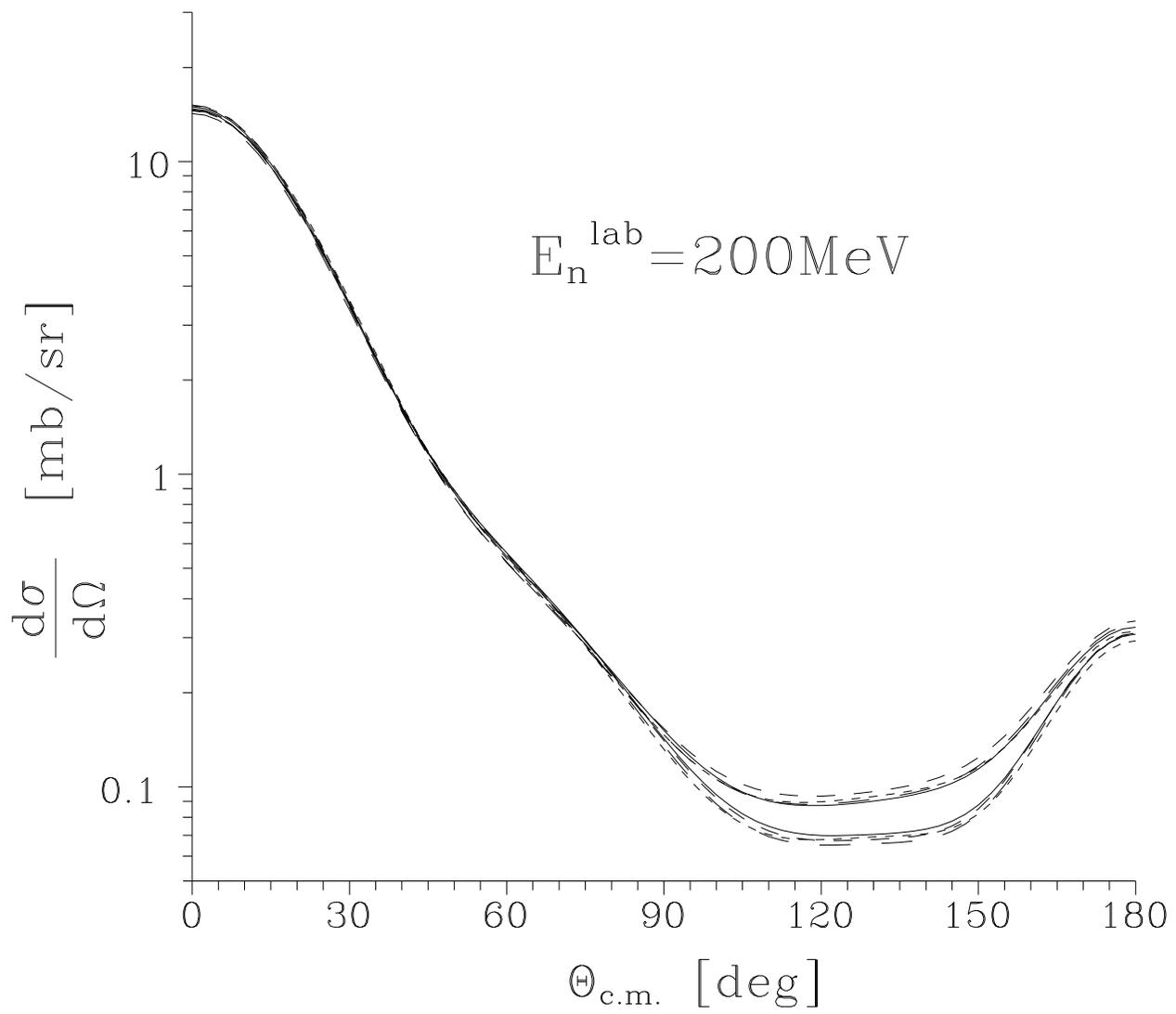
\end{document}